\documentclass[12pt,epsfig,rotating] {article}
\usepackage[dvips]{graphics}
\begin{document}
\newcommand{\sqrts}{\mbox{$\protect \sqrt{s}$}}
\newcommand{\sqrtsp}{\mbox{$\sqrt{s'}$}}
\newcommand{\epm} {\mbox{$\mathrm{e}^+ \mathrm{e}^-$}}
\newcommand{\mpm} {\mbox{$\mu^+ \mu^-$}}
\newcommand{\nprode}{\mbox{$N^{\mathrm{e}^+ \mathrm{e}^-}_{prod}$}}
\newcommand{\nprodm}{\mbox{$N^{\mu^+ \mu^-}_{prod}$}}
\newcommand{\nexpe}{\mbox{$N_{exp}^{\mathrm{e}^+ \mathrm{e}^-}$}}
\newcommand{\nexpm}{\mbox{$N_{exp}^{\mu^+ \mu^-}$}}
\newcommand{\nexpt}{\mbox{$N_{exp}^{total}$}}
\newcommand{\Zo}{\mbox{$\protect {\rm Z}$}}
\newcommand{\bZo}{{\bf \mbox{$\rm Z}^0$}}
\newcommand{\Zs}{\mbox{$\mathrm{Z}^{*}$}}
\newcommand{\ho}{\mbox{$\mathrm{h}$}}
\newcommand{\Ho}{\mbox{$\mathrm{H}^{0}$}}
\newcommand{\Ao}{\mbox{$\mathrm{A}$}}
\newcommand{\Wpm}{\mbox{$\mathrm{W}^{\pm}$}}
\newcommand{\Hpm}{\mbox{$\mathrm{H}^{\pm}$}}
\newcommand{\WW}{\mbox{$\mathrm{W}^{+}\mathrm{W}^{-}$}}
\newcommand{\ZZ}{\mbox{$\mathrm{Z}^{0}\mathrm{Z}^{0}$}}
\newcommand{\ko}{\mbox{${\tilde{\chi_1}^0}$}}
\newcommand{\koko}{\mbox{${\tilde{\chi_1}^0}{{\tilde{\chi_1}^0}}$}}
\newcommand{\bho}{\mbox{$\boldmath{\mathrm{H}^{0}}$}}
\newcommand{\ee}{\mbox{$\mathrm{e}^{+}\mathrm{e}^{-}$}}
\newcommand{\bee}{\mbox{\boldmath{$\mathrm{e}^{+}\mathrm{e}^{-}$}\unboldmath}}
\newcommand{\mm}{\mbox{$\mu^{+}\mu^{-}$}}
\newcommand{\bmm}{\mbox{$\boldmath {\mu^{+}\mu^{-}} $}}
\newcommand{\nn}{\mbox{$\nu \bar{\nu}$}}
\newcommand{\bnn}{\mbox{$\boldmath {\nu \bar{\nu}} $}}
\newcommand{\qq}{\mbox{$\protect {\rm q} \protect \bar{{\rm q}}$}}
\newcommand{\ff}{\mbox{$\mathrm{f} \bar{\mathrm{f}}$}}
\newcommand{\bqq}{\mbox{$\boldmath {\mathrm{q} \bar{\mathrm{q}}} $}}
\newcommand{\pb}{\mbox{$\protect {\rm pb}^{-1}$}}
\newcommand{\ra}{\mbox{$\rightarrow$}}
\newcommand{\br}{\mbox{$\boldmath {\rightarrow}$}}
\newcommand{\erh}{\mbox{$\mathrm{e}^+\mathrm{e}^-\rightarrow\mathrm{hadrons}$}}
\newcommand{\tptm}{\mbox{$\tau^{+}\tau^{-}$}}
\newcommand{\ftptm}{\mbox{\boldmath{$\tau^{+}\tau^{-}$}\unboldmath}}
\newcommand{\tpm}{\mbox{$\tau^{\pm}$}}
\newcommand{\pzvis}{\mbox{$\protect P^z_{\rm vis}$}}
\newcommand{\evisn}{\mbox{$\protect E_{\rm vis}$/$\protect \sqrt{s}$}}
\newcommand{\gamgam}{\mbox{$\gamma\gamma$}}
\newcommand{\uu}{\mbox{$\mathrm{u} \bar{\mathrm{u}}$}}
\newcommand{\dd}{\mbox{$\mathrm{d} \bar{\mathrm{d}}$}}
\newcommand{\bb}{\mbox{$\mathrm{b} \bar{\mathrm{b}}$}}
\newcommand{\fbb}{\mbox{\boldmath{$\mathrm{b} \bar{\mathrm{b}}$}\unboldmath}}
\newcommand{\cc}{\mbox{$\mathrm{c} \bar{\mathrm{c}}$}}
\newcommand{\nunu}{\mbox{$\nu \bar{\nu}$}}
\newcommand{\mZ}{\mbox{$m_{\mathrm{Z}}$}}
\newcommand{\mH}{\mbox{$m_{\mathrm{H}^{0}}$}}
\newcommand{\mh}{\mbox{$m_{\mathrm{h}}$}}
\newcommand{\mA}{\mbox{$m_{\mathrm{A}}$}}
\newcommand{\mah}{\mbox{$m_{\mathrm{h}/\mathrm{A}}$}}
\newcommand{\mHpm}{\mbox{$m_{\mathrm{H}^{\pm}}$}}
\newcommand{\xiu}{\mbox{$\xi_u$}}
\newcommand{\xid}{\mbox{$\xi_d$}}
\newcommand{\xad}{\mbox{$\xi^{\mathrm{A}}_d$}}
\newcommand{\xhd}{\mbox{$\xi^{\mathrm{h}}_d$}}
\newcommand{\xau}{\mbox{$\xi^{\mathrm{A}}_u$}}
\newcommand{\xhu}{\mbox{$\xi^{\mathrm{h}}_u$}}
\newcommand{\xahd}{\mbox{$\xi^{\mathrm{h}/\mathrm{A}}_d$}}
\newcommand{\xahu}{\mbox{$\xi^{\mathrm{h}/\mathrm{A}}_u$}}
\newcommand{\fxah}{\mbox{\boldmath{$\xi^{\mathrm{h}/\mathrm{A}}$}\unboldmath}}
\newcommand{\xnfa}{\mbox{$\xi_{d}^{95}$}}
\newcommand{\xnfh}{\mbox{$\xi_{d}^{95}$}}
\newcommand{\kao}{\mbox{$\kappa_{\mathrm{A}}$}}
\newcommand{\kho}{\mbox{$\kappa_{\mathrm{h}}$}}
\newcommand{\mW}{\mbox{$m_{\mathrm{W}^{\pm}}$}}
\newcommand{\mtop}{\mbox{$m_{\mathrm{t}}$}}
\newcommand{\mb}{\mbox{$m_{\mathrm{b}}$}}
\newcommand{\lpm}{\mbox{$\ell ^+ \ell^-$}}
\newcommand{\G}{\mbox{$\mathrm{GeV}$}}
\newcommand{\Gc}{\mbox{$\mathrm{GeV}$}}
\newcommand{\Gcs}{\mbox{$\mathrm{GeV}$}}
\newcommand{\epsnn}{\mbox{$\epsilon^{\nu\bar{\nu}}$(\%)}}
\newcommand{\Nnn}{\mbox{$N^{\nu \bar{\nu}}_{exp}$}}
\newcommand{\epsll}{\mbox{$\epsilon^{\ell^{+}\ell^{-}}$(\%)}}
\newcommand{\Nll}{\mbox{$N^{\ell^+\ell^-}_{exp}$}}
\newcommand{\Nexp}{\mbox{$N^{total}_{exp}$}}
\newcommand{\kl}{\mbox{$\mathrm{K_{L}}$}}
\newcommand{\dedx}{\mbox{d$E$/d$x$}}
\newcommand{\etal}{\mbox{$et$ $al.$}}
\newcommand{\ie}{\mbox{$i.e.$}}
\newcommand{\sba}{\mbox{$\sin ^2 (\beta -\alpha)$}}
\newcommand{\cba}{\mbox{$\cos ^2 (\beta -\alpha)$}}
\newcommand{\tanb}{\mbox{$\tan \beta$}}
\newcommand{\PhysLett}  {Phys.~Lett.}
\newcommand{\PRL} {Phys.~Rev.\ Lett.}
\newcommand{\PhysRep}   {Phys.~Rep.}
\newcommand{\PhysRev}   {Phys.~Rev.}
\newcommand{\PRD}[3]   {Phys.~Rev.\ {\bf{D#1}} (#2) #3}
\newcommand{\NPhys}  {Nucl.~Phys.}
\newcommand{\NIM} {Nucl.~Instrum.\ Meth.}
\newcommand{\NIMA}[1]   {\NIM\ {\bf A{#1}}}
\newcommand{\CPC}[1]      {Comp.\ Phys.\ Comm.\ {\bf #1}}
\newcommand{\ZPhys}  {Z.~Phys.}
\newcommand{\ZPhysC}[1] {Z. Phys. {\bf C#1}}
\newcommand{\IEEENS} {IEEE Trans.\ Nucl.~Sci.}
\newcommand{\thelimit}  {82}
\newcommand{\ZPC}[3] {Z.~Phys.\ {\bf{C#1}} (#2) #3}
\newcommand{\PLB}[3] {Phys.~Lett.\ {\bf{B#1}} (#2) #3}
\newcommand{\APP}[3] {Acta Phys.~Polon.\ {\bf{B#1}} (#2) #3}
\renewcommand{\topfraction}{0.99}    
\renewcommand{\bottomfraction}{0.99} 

\begin{titlepage}
\centerline{\Large EUROPEAN ORGANIZATION FOR NUCLEAR RESEARCH}
\begin{flushright}
      CERN-EP-2001-077   \\ 01 November 2001
\end{flushright}
\bigskip\bigskip\bigskip

\begin{center}{\LARGE\bf   Search for Yukawa Production of a \\[0.3cm]
Light Neutral Higgs Boson at LEP}

\end{center}
 \bigskip
\begin{center}{\LARGE The OPAL Collaboration}
\end{center}
\begin{center}{\large  Abstract}\end{center}
Within a Two-Higgs-Doublet Model (2HDM) a search for a light Higgs
boson in the mass range of 4--12 GeV has been performed in the Yukawa
process \epm $\rightarrow$ \bb\Ao/\ho\ $\rightarrow$ \bb\tptm, using
the data collected by the OPAL detector at LEP between 1992 and 1995
in \epm\ collisions at about 91 GeV centre-of-mass energy.  A
likelihood selection is applied to separate background and signal.
The number of observed events is in good agreement with the expected
background. Within a CP-conserving 2HDM type II model the
cross-section for Yukawa production depends on $\xad=|\tan\beta|$ and
$\xhd=|\sin\alpha/\cos\beta|$ for the production of the CP-odd \Ao\
and the CP-even \ho, respectively, where $\tan\beta$ is the ratio of
the vacuum expectation values of the Higgs doublets and $\alpha$ is
the mixing angle between the neutral CP-even Higgs bosons. From our
data 95\% C.L.  upper limits are derived for $\xad$ within the range
of 8.5 to 13.6 and for $\xhd$ between 8.2 to 13.7, depending on the
mass of the Higgs boson, assuming a branching fraction into \tptm\ of
100\%. An interpretation of the limits within a 2HDM type II model
with Standard Model particle content is given.  These results impose
constraints on several models that have been proposed to explain
the recent BNL measurement of the muon anomalous magnetic moment.
\bigskip
\begin{center}
{\bf (To be submitted to Eur. Phys. J. C)}
\end{center}
\bigskip\bigskip\bigskip\bigskip

\end{titlepage}


\begin{center}{\Large        The OPAL Collaboration
}\end{center}\bigskip
\begin{center}{
G.\thinspace Abbiendi$^{  2}$,
C.\thinspace Ainsley$^{  5}$,
P.F.\thinspace {\AA}kesson$^{  3}$,
G.\thinspace Alexander$^{ 22}$,
J.\thinspace Allison$^{ 16}$,
G.\thinspace Anagnostou$^{  1}$,
K.J.\thinspace Anderson$^{  9}$,
S.\thinspace Arcelli$^{ 17}$,
S.\thinspace Asai$^{ 23}$,
D.\thinspace Axen$^{ 27}$
% ...
}\end{center}\bigskip
\bigskip
$^{  1}$School of Physics and Astronomy, University of Birmingham,
Birmingham B15 2TT, UK
\newline
$^{  2}$Dipartimento di Fisica dell' Universit\`a di Bologna and INFN,
I-40126 Bologna, Italy
\newline
$^{  3}$Physikalisches Institut, Universit\"at Bonn,
D-53115 Bonn, Germany
\newline
% ...
$^{  5}$Cavendish Laboratory, Cambridge CB3 0HE, UK
\newline
% ...
$^{  9}$Enrico Fermi Institute and Department of Physics,
University of Chicago, Chicago IL 60637, USA
\newline
% ...
$^{ 16}$Department of Physics, Schuster Laboratory, The University,
Manchester M13 9PL, UK
\newline
$^{ 17}$Department of Physics, University of Maryland,
College Park, MD 20742, USA
\newline
% ...
$^{ 22}$Department of Physics and Astronomy, Tel Aviv University,
Tel Aviv 69978, Israel
\newline
$^{ 23}$International Centre for Elementary Particle Physics and
Department of Physics, University of Tokyo, Tokyo 113-0033, and
Kobe University, Kobe 657-8501, Japan
\newline
% ...
$^{ 27}$University of British Columbia, Department of Physics,
Vancouver BC V6T 1Z1, Canada
\newline
% ...
$^{  b}$ and Royal Society University Research Fellow
\newline
$^{  c}$ and Institute of Nuclear Research, Debrecen, Hungary
\newline
$^{  e}$ and Heisenberg Fellow
\newline
$^{  f}$ and Department of Experimental Physics, Lajos Kossuth University,
 Debrecen, Hungary
\newline
% ...
$^{  i}$ and Research Institute for Particle and Nuclear Physics,
Budapest, Hungary
\newline
% ...

\bigskip

\cleardoublepage

\section{Introduction}
The search for the last missing particle predicted by the Standard
Model, the Higgs boson, is one of the main topics in high energy
physics at LEP. The Standard Model (SM) has been verified to a very
high degree of precision but no Higgs boson has yet been
discovered. Current searches at LEP in the Standard Model
scenario~\cite{LEPSM} exclude Higgs bosons with masses $\mH$ below
114.1 GeV at the 95\%~confidence level. Many proposed models extend
the SM while preserving the good agreement with experimental data. A
minimal extension of the SM, the Two-Higgs-Doublet Model (2HDM), has
this property.  In non-supersymmetric 2HDMs, Higgs bosons with small
masses still cannot be excluded~\cite{Where,CKZ}.  The analysis
presented here, a search for \Ao\ and \ho\ in the mass range 4--12
GeV, provides new constraints on the parameter space of these models.

In the 2HDM, two complex Higgs doublets are introduced 
to generate the mass of the fermions:
\begin{equation}
\Phi_1={\phi_{1}^{+} \choose \phi_{1}^{0}}
\qquad\mbox{;}\qquad
\Phi_2={\phi_{2}^{+} \choose \phi_{2}^{0}}.
\end{equation}
One constraint for extended Higgs sectors is the experimental observation
that the value of $\rho\equiv\mW^2/(\mZ^2\cos^2\theta_{\rm W})
\approx 1$.  The condition of $\rho\approx 1$ is automatically met by models
with only Higgs doublets.

There are several possible patterns for how the new fields may couple
to fermions.  Four types of 2HDMs are theoretically considered in
order to avoid introducing flavour changing neutral currents (FCNC)
\cite{Hunter}. The four types differ in the way the two Higgs fields
$\Phi_{1}$ and $\Phi_{2}$ couple to fermions (see
Table~\ref{masscoup}).

\begin{table}[htb]
  \begin{center}
    \begin{tabular}{|c|c|c|c|c|}\hline
      couples to & type I & type II & type III & type IV \\ \hline
      down-type leptons &$\Phi_{2}$ & $\Phi_{1}$&$\Phi_{2}$&$\Phi_{1}$\\ \hline
      up-type quarks & $\Phi_{2}$& $\Phi_{2}$&$\Phi_{2}$&$\Phi_{2}$\\ \hline
      down-type quarks & $\Phi_{2}$& $\Phi_{1}$&$\Phi_{1}$&$\Phi_{2}$\\ \hline
    \end{tabular}
  \end{center}
  \caption{{\it The couplings of the Higgs fields according to the four types
      of 2HDMs.}}
  \label{masscoup}
\end{table}

In this analysis the CP-conserving 2HDM type II is considered where
$\Phi_{1}$ couples to the down-type and the $\Phi_{2}$ couples to the
up-type matter fields.  This 2HDM predicts five physical Higgs bosons:
two charged (\Hpm), two CP-even (\Ho, \ho) and one CP-odd (\Ao) Higgs
boson (the neutral Higgs bosons are often referred to as {\it scalar}
and {\it pseudoscalar}, respectively).  The Higgs sector of the
general CP-conserving 2HDM has six free physical
parameters~\cite{Hunter} and can be parametrized by the masses
$\mHpm$, $\mH$, $\mh$ and $\mA$ and two dimensionless parameters
$\alpha$ and $\tan\beta$, with $\alpha$ being the mixing angle in the
CP-even neutral Higgs sector, and $\tan\beta=v_2/v_1$ being the ratio
of the two vacuum expectation values $v_2$ and $v_1$.

The Higgs sector in the Minimal Supersymmetric Standard Model (MSSM)
is a special case of a 2HDM type II model in which, due to
relations imposed by supersymmetry, only two free parameters remain
(eg. $\tan\beta$ and $\alpha$) at tree level in the Higgs sector.
Direct searches at LEP in the MSSM scenario have set limits on the
Higgs mass and its parameter space.  For example in a parameter scan
where the parameter combination was chosen to give a maximal predicted
mass $\mh$, 95\%\ C.L. limits have been set on the masses $\mh$ and
$\mA$ larger than 88.4 GeV for $\tan\beta > 0.4$~\cite{LEPHiggs}.
  
\section{Higgs Boson Production in \bee\ Collisions}
\begin{figure}[h]
  \begin{center}
    \resizebox{160mm}{!}{
      \includegraphics{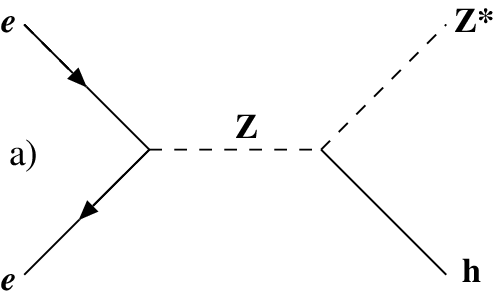}\hspace{0.5cm}
      \includegraphics{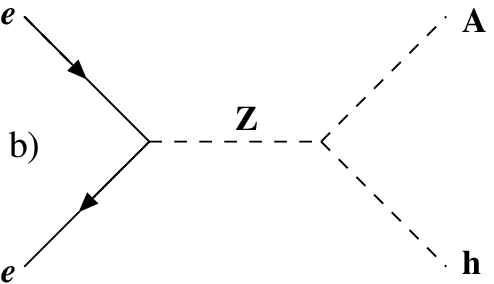}\hspace{0.5cm}
      \includegraphics{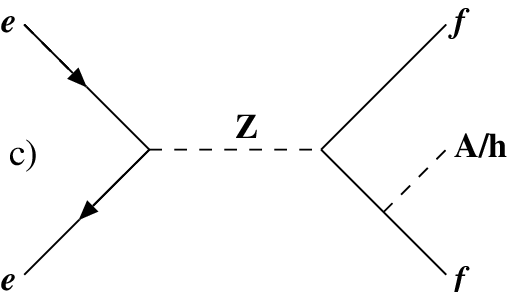}}
  \end{center}
  \caption{{\it The three Born level Higgs production processes in a 
        2HDM at LEP: The Higgsstrahlung process a), associated production b) 
        and the Yukawa process c).}}
  \label{alleprozesse}
\end{figure}

There are three main processes for Higgs production at Born level
within the energy range covered by LEP, namely the Higgsstrahlung,
associated production and Yukawa processes shown in Figure
\ref{alleprozesse}, of which only the first process is of importance
in the Standard Model.  In the Standard Model the Yukawa process is
suppressed by the factor $(m_{\rm f}^2/\mh^2)$, and the associated
production of Higgs bosons is nonexistent.
The cross-sections of the 2HDM are closely related to similar processes of
the Standard Model: 
\begin{equation}
\sigma (\epm \rightarrow \Zo \rightarrow \ho{\rm Z}^{*}) = \sigma_{\rm SM} 
(\epm \rightarrow \Zo \rightarrow {\rm H}_{\rm SM}{\rm Z}^{*})
\sin^2 (\beta - \alpha)
\end{equation}
\begin{equation}
\sigma (\epm \rightarrow \Zo \rightarrow \ho\Ao) = \sigma_{\rm SM} 
(\epm \rightarrow \Zo \rightarrow \nu\bar{\nu})
\cos^2 (\beta - \alpha)\lambda^{\frac{3}{2}}
\end{equation}
where $\lambda=(1-\kho-\kao)^2
-4\kho^2\kao^2$ being a 
phase space factor, with $\kappa_i=m_i^2/\mZ^2$ \cite{Hunter}.

In recent years there have been searches for the Standard Model Higgs
boson as well as for MSSM Higgs bosons by all four LEP experiments
\cite{LEPHiggs}.  The interpretation of the flavour independent Higgs
search within the 2HDM in the mass range below approximately $40$ GeV
requires \sba\ to be less than 0.2 \cite{LimitAB}.  For a sufficiently
small \sba, the \ho\ produced through the Higgsstrahlung process can
not be seen in the data collected at LEP due to the cross-section
suppression factor (Eqn. 2).  Associated production, if kinematically
allowed, would be the dominant process for Higgs boson production. On
the other hand a light Higgs boson which is produced only in the
Yukawa process (\Ao\ or \ho\ whichever is lighter) could have escaped
discovery.  Under the assumption that the Higgsstrahlung process is
suppressed and associated production is kinematically forbidden,
$\mA+\mh >\sqrt{s}$, the Yukawa process becomes the dominant process
for Higgs production at LEP. This scenario can easily be realised in
the general 2HDM since its parameters are not constrained
theoretically \cite{Where,CKZ}.  In this analysis we concentrate on
Higgs masses below the 2$m_{\rm b}$ threshold since previous analyses
\cite{OPALPaper} are insensitive to such light Higgs bosons.  We will
also constrain ourselves to Higgs masses above the 2$m_{\tau}$
threshold.
 
The cross-section of the Yukawa process \cite{Maria1}
\begin{equation}
\sigma_{\rm Yukawa}\propto m_{\mathrm f}^2 \, N_{\mathrm c} \, 
\xi_{\mathrm f}^2
\end{equation}
is proportional to the squared fermion mass, $m_{\mathrm f}^2$, the
colour factor, $N_{\mathrm c}$, of the emitting fermion, and an
enhancement factor $\xi_{\mathrm f}^2$, which describes the
coupling between the Higgs boson and the emitting fermion (see Table
\ref{tab1}).

\begin{table}[htb]
\begin{center}
\begin{tabular}{|c|l|l|}\hline
 Higgs Type  & Down Type Fermions& Up Type Fermions\\ \hline
  \Ao\ & $\xad=\tan\beta$ & $\xau=1/\tan\beta$ \\ 
  \ho\ & $\xhd=-\sin\alpha/\cos\beta$ & $\xhu=\cos\alpha/\sin\beta$ \\ \hline
\end{tabular}
\caption{{\it The enhancement factor $\xi_{\mathrm f}$ depending on
the type of the Higgs boson and the emitting fermion.}}
\label{tab1}
\end{center}
\end{table}

The range of $\xahd$ for which a detectable signal would be
produced can be divided into two parts:
\begin{enumerate}
\item $\xid<1$ (implies $\xiu>1$): An up-type quark pair \cc\ radiates a
Higgs boson decaying into \cc.
\item $\xid>1$: A down-type quark pair \bb\ radiates a Higgs boson
  decaying into \tptm.
\end{enumerate}

\begin{figure}[htb]
  \begin{center}
    \resizebox{120mm}{!}{
      \includegraphics{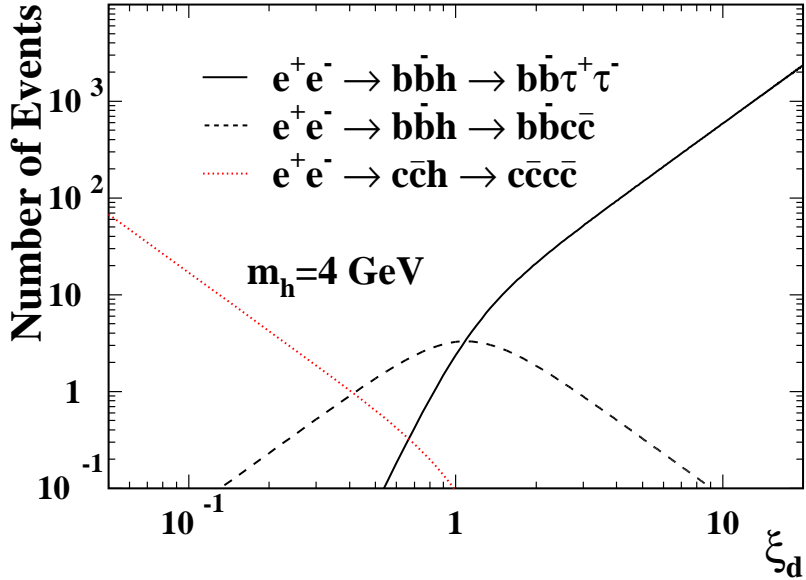}}
  \caption{{\it The number of expected signal events for a scalar
      Higgs \ho\ with $\mh$=4 GeV calculated for a luminosity of 113.1
      \pb\ at $\sqrt{s}=$\mZ.}}
    \label{numevt}
  \end{center}
\end{figure}

In this analysis, only the \bb\tptm\ final state is considered, since
background suppression can be performed more efficiently using the
clear signature of bottom decay, together with the missing energy and
low-multiplicity signature of $\tau$ decays. In Figure \ref{numevt}
the respective numbers of expected events are shown as a function of
$\xhd$ for the example of a CP-even Higgs with a mass of 4~GeV.

\section{Search in the \fbb\ftptm\ Channel}
\subsection{Data and Monte Carlo Samples}
The present analysis is based on data collected with the OPAL detector
\cite{OPALDetector} during the years 1992--1995, taken at
centre-of-mass energies close to the \Zo\ peak. Although the integrated
luminosity at LEP recorded at $\sqrt{s}$ between 130 and 208 GeV is
about a factor five higher than the luminosity around the \Zo\ peak,
the number of produced b quarks is about a factor of 100 smaller in
the higher-energy data.  Therefore we only use data recorded at the
\Zo\ peak at centre-of-mass energies near 91 GeV to search for Yukawa
production of Higgs bosons.
The data collected at off-peak energies in the range between 89 GeV
and 93 GeV are included to increase the available statistics.  Since
the characteristics of signal events do not depend on the precise
center-of-mass energy, and since the background is completely
dominated by SM hadronic \Zo\ decays, we can treat the data as if all
were taken on-peak by rescaling the off-peak luminosity appropriately.
For this we count the number of observed hadronic events~\cite{OPALZ1}
in the data.  This number, efficiency corrected and divided by the peak
cross-section for hadronic \Zo\ decays of 30.45 nb as measured by
OPAL~\cite{Lumi}, then yields an effective on-peak luminosity of 113.1
\pb~\cite{OPALZ2}, from which about 12\%\ is contributed by the data
taken off-peak.

We consider three types of background classes, with the full
detector response simulated as described in~\cite{osim}, using the
following generators.

\begin{enumerate}
\item {\bf Two-photon background}, generated with Vermaseren 
  1.01~\cite{Vermaseren} and PHOJET~\cite{PHOJET}

\begin{itemize} 
\item[] Two-photon production of hadronic final states is
    characterized by little visible energy in the detector. Usually, the
    e$^+$ and e$^-$ escape undetected close to the beam axis, causing only
    a small amount of transverse momentum. The relatively large
    cross-section of this process nevertheless makes us consider it as a
    potential background.  The Monte Carlo sample of about two million
    two-photon events corresponds to a luminosity of about four times the
    data luminosity.
\end{itemize}
\item {\bf Four-fermion background}, generated with FERMISV~\cite{Fermisv} 
  and grc4f~\cite{grc4f}
  \begin{itemize}
  \item[] The four-fermion background at LEP 1 mainly originates from
    Initial State Pair (ISP) and Final State Pair (FSP) radiation
    diagrams.  This background class can be divided into two
    subclasses, the first containing four lepton final states and the
    second having \qq\ff\ final states. The first subclass is
    eliminated by requiring the event to pass a general multihadronic
    selection~\cite{OPALZ2}.  The second class contains a small amount
    of irreducible \epm $\rightarrow$ \bb\lpm\ background with a
    charged lepton pair, mainly from FSP radiation. The Monte Carlo
    sample of about 25000 four-fermion events corresponds to a
    luminosity of about eight times the data luminosity.
  \end{itemize}
\item {\bf \boldmath \qq\ background\unboldmath}, generated with Jetset
  7.4~\cite{Jetset}
  \begin{itemize}
  \item[] This class of background consists of events of the type \epm
    $\rightarrow \Zo / \gamma \rightarrow \qq$.  Events with gluons
    radiated off the quarks, especially \epm $\rightarrow \Zo / \gamma
    \rightarrow \bb {\rm g(g)}$, are very likely to have signal
    characteristics and represent the main background in this
    analysis. The Monte Carlo sample of about seven million processed
    \qq\ events corresponds to a luminosity about two times the data
    luminosity.  Generated Monte Carlo \qq\ background samples for
    different detector setups are weighted according to the respective
    luminosity for the data.
  \end{itemize} 
\end{enumerate}

Eighteen signal samples of 10000 events each with masses of
$\mah$=4--12 GeV in one GeV steps were generated using a newly written
Monte Carlo program based on \cite{Maria1}.  The hadronisation is done
with JETSET version 7.408 \cite{Jetset} together with OPAL specific
modifications~\cite{OPMOD}. The decay of the tau leptons is simulated
with the tau decay library TAUOLA~\cite{Tauola}.  These signal
events are subjected to the same reconstruction and event selection as
the real data.

\subsection{Analysis Tools}
In calculating the visible energies and momenta, $E_{vis}$ and
$\vec{P}_{vis}$, of individual jets and of the total event,
``energy-flow objects'' are formed from the charged tracks and
calorimeter clusters \cite{MT}.  To avoid double counting, the
energies expected to be deposited in the calorimeters by the charged
particles are subtracted from the energies of the associated
calorimeter clusters.

In order to identify jets containing b hadron decays, three
independent techniques using lifetime, high-$p_t$ lepton
characteristics and kinematic information are used.  Artificial Neural
Networks (ANN's) have been trained to combine several
lifetime-sensitive tagging variables and kinematic variables. For each
jet, the outputs of the lifetime ANN, the kinematic ANN, and the
lepton tag are combined into a likelihood variable $B$ which
discriminates b-flavoured jets from c-flavoured jets and light quark
jets~\cite{OPALPaper}.

\subsection{Properties of Yukawa Production}
\begin{figure}[htb]
  \begin{center} 
    \resizebox{140mm}{!}{\includegraphics{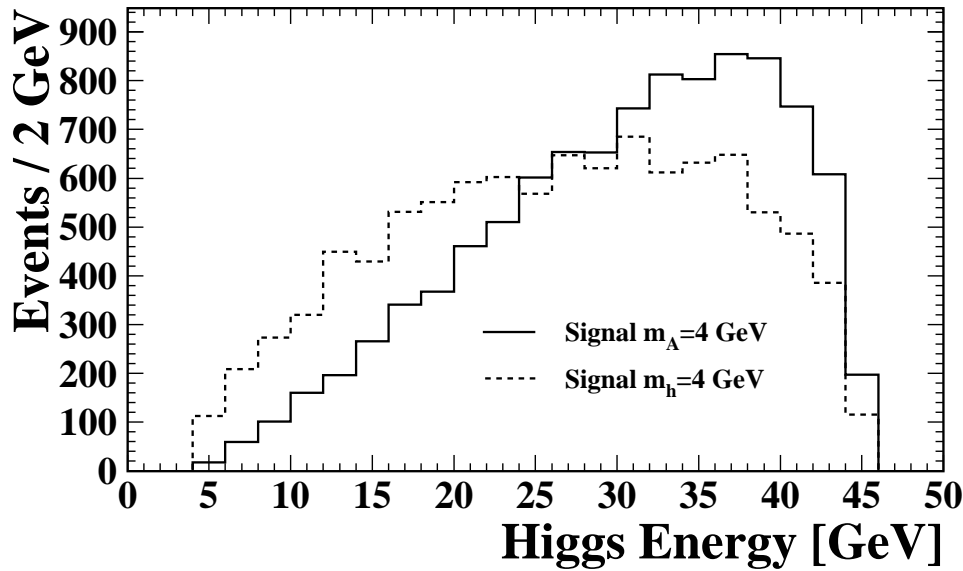}} 
    \caption{{\it The energy distribution of the Higgs boson with
       $\mah=4$ {\rm GeV} from 10000 Monte Carlo signal events at generator
       level.}}  \label{vergleichah}
  \end{center}
\end{figure}

\begin{figure}[htb]
  \begin{center}
    \resizebox{100mm}{!}{
      \includegraphics{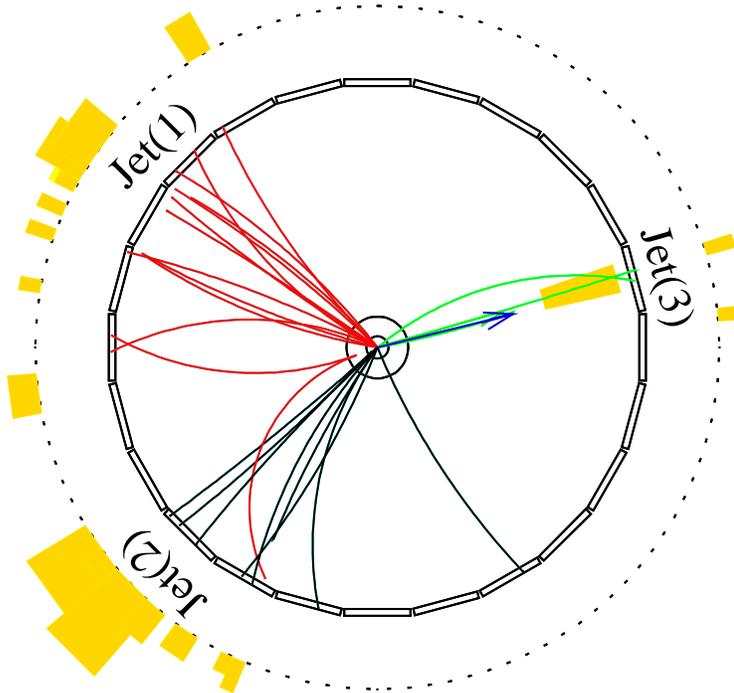}}
  \caption{{\it A characteristic simulated signal event \epm $\rightarrow$
      \bb${\Ao}/{\ho} \rightarrow$ \bb\tptm reconstructed in the OPAL
      detector. Jet(1) and Jet(2) contain the tracks
      of the hadronized {\rm b} quarks and Jet(3) contains the decay
      products of the Higgs boson.  The missing momentum vector (dark
      grey arrow) points along the Jet(3) axis due to the
      undetected neutrinos in the decay of the $\tau$'s.}}
    \label{event8699.22}
  \end{center}
\end{figure}

One of the properties of Yukawa production is the hard energy spectrum
of the emitted Higgs bosons. In Figure \ref{vergleichah} the energy
distributions of the \Ao\ and \ho\ are shown. This leads to a high
boost of the decay products of the Higgs particles and results in a
small angle between the two $\tau$'s in the detector.  Consequently,
the tracks of the decay products of the two $\tau$'s can be
reconstructed in one single low-multiplicity jet. In addition, two
high-multiplicity jets are expected to be associated with the b
quarks. Thus the expected topology of a signal event has a three jet
signature.  A simulated signal event is shown in Figure
\ref{event8699.22}.  In this analysis each event is forced into three
jets using the Durham algorithm \cite{Durham}. The jets are sorted
according to their multiplicity, assuming that the jet with the lowest
charged particle multiplicity (denoted `Jet(3)') contains the two
$\tau$ leptons from the Higgs decay.

\subsection{Event Selection}
The event selection consists of two parts, a preselection and a
subsequent likelihood selection.  Since the unknown Higgs mass enters
in the properties of the likelihood variables, the selection was
performed and optimized separately for each of the nine simulated
signal mass hypotheses.

\begin{figure}[htb]
  \begin{center}
    \resizebox{160mm}{!}{
      \includegraphics{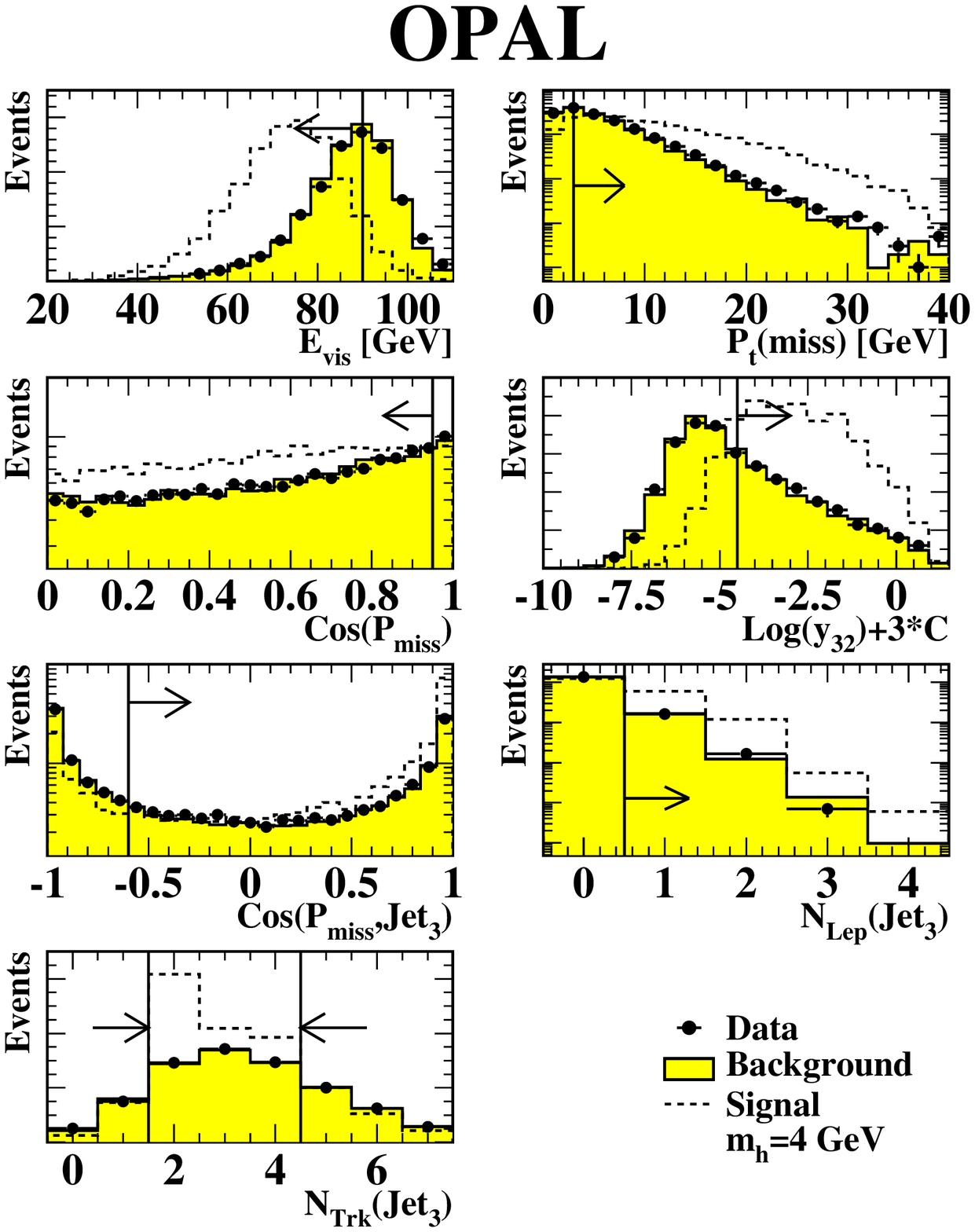}}
  \caption{{\it The preselection variables and their impact on processed
      data. The points with error bars are data, while the solid histogram is
      the simulation of the \qq\ background normalised to the 
      recorded luminosity.
      The dashed line represents a simulated signal of a scalar Higgs 
      with $\mh=4$ {\rm GeV} scaled arbitrarily for better 
      visibility. 
      The arrows indicate the cuts made on the variables.}}
  \label{prevars}
\end{center}
\end{figure}

The preselection consists of the eight cuts described below (see
Figure \ref{prevars}).
 
\begin{itemize}
\item[0.] General hadronic event selection as described in~\cite{OPALZ2}.

\item[1.] $E_{vis} < 90$ GeV. Due to the neutrinos from $\tau$
  decays the signal events have missing energy in the detector.  This
  cut on the visible energy primarily suppresses \qq\ background.
  
\item[2.] $P_t(miss) > 3$ GeV. The cut on the transverse missing
  momentum in the event is additionally introduced to reduce
  two-photon background.
  
\item[3.] To suppress two-photon events further, we require the
missing momentum vector to have $|\cos(\rm{beam~axis},\vec{P}_{miss})|
< 0.95$.

  \item[4.] A two dimensional cut $3\cdot C + \log(y_{32}) \ge -4.5$
    on the event shape variables, $C$-value \cite{CVAR} and $y_{32}$
    \cite{Y43}, is introduced to suppress two-jet like events.

  \item[5.] A large fraction of the missing momentum in the event is due
    to the undetectable neutrinos of the decaying $\tau$'s in Jet(3). 
    Therefore a cut on the angle between Jet(3) and $\vec{P}_{miss}$ is
    introduced: $\cos(\mbox{Jet(3)},\vec{P}_{miss}) > -0.6$.
    
  \item [6.] We require at least one identified electron or muon in
    Jet(3).  Electrons are identified according to \cite{elecid} and
    muons according to \cite{muonid}.  This cut is made to reduce
    \qq\ background and to define efficient likelihood variables.  The
    probability of two $\tau$ decays containing at least one charged
    lepton is approximately 60\%. To ensure the correct efficiency
    modelling in the Monte Carlo a further cut on the lepton momentum,
    $P_l > 2$~GeV, is added.
    
  \item [7.] 2 $\le N_{track} \le$ 4 in Jet(3). This cut on the
    charged multiplicity of Jet(3) is introduced in order to optimize
    the ratio of signal over square root of background after
    the likelihood selection.
\end{itemize}

As shown in Table \ref{cutflow}, the observed number of data
events and the expected background agree well at each step of the
preselection. Of the background passing the general multihadronic
selection (cut 0) about 2\% of the four-fermion events, 0.4\% of the
\qq\ events and 0.3\% of the two-photon events remain after the
complete preselection, with the \qq\ background clearly being the
dominant contribution.  The signal selection efficiency, only weakly
dependent of the Higgs boson mass, ranges from 11\% to 17\%, depending on
the signal sample, as detailed for some typical masses in
Table~\ref{cuteff_a}.

\begin{table}
  \begin{center}
    \begin{tabular}{|c|r|r|r|r|} \hline
      Cuts & Data & \qq & four-fermion & two-photon \\ \hline 
      (0) &  338 $\times 10^4$&  338 $\times 10^4$& 887 & 346 \\
      (1) &  196 $\times 10^4$&  198 $\times 10^4$& 545 & 236 \\
      (2) &  131 $\times 10^4$&  128 $\times 10^4$& 357 & 123 \\
      (3) &  125 $\times 10^4$&  123 $\times 10^4$& 330 &  54 \\
      (4) &  622 $\times 10^3$&  599 $\times 10^3$& 198 &  43 \\
      (5) &  400 $\times 10^3$&  383 $\times 10^3$& 125 &  32 \\
      (6) &  292 $\times 10^2$&  307 $\times 10^2$&  39 &   5 \\
      (7) &  142 $\times 10^2$&  141 $\times 10^2$&  22 &   1 \\ \hline 
     \end{tabular}
     \caption{ {\it The number of events selected and expected in the
         preselection. The three categories of background have been
         normalised to the data luminosity.}}  \label{cutflow}
         \end{center}
\end{table}

\begin{table}[htb]
  \begin{center}
    \begin{tabular}{|c|c|c|c|c|} \hline
      Cuts & Efficiency (\%) & Efficiency (\%)& Efficiency (\%)& Efficiency (\%) \\  
      &\mA=4 GeV &\mA=10 GeV & \mh=4 GeV&\mh=10 GeV \\ \hline 
      (1) & 92.0 & 92.7 & 90.2 & 91.6 \\
      (2) & 80.6 & 83.0 & 76.7 & 81.4 \\
      (3) & 78.7 & 80.8 & 74.9 & 79.4 \\
      (4) & 64.5 & 73.9 & 54.9 & 71.2 \\
      (5) & 54.8 & 65.4 & 43.8 & 60.7 \\
      (6) & 19.2 & 22.2 & 14.2 & 19.3 \\
      (7) & 16.2 & 17.4 & 11.2 & 14.7 \\ \hline 
     \end{tabular}
     \caption{ {\it The efficiency of the preselection for selected Higgs 
         masses.}}
  \label{cuteff_a}
  \end{center}
\end{table}

\begin{figure}[htb]
  \begin{center}
    \resizebox{160mm}{!}{
      \includegraphics{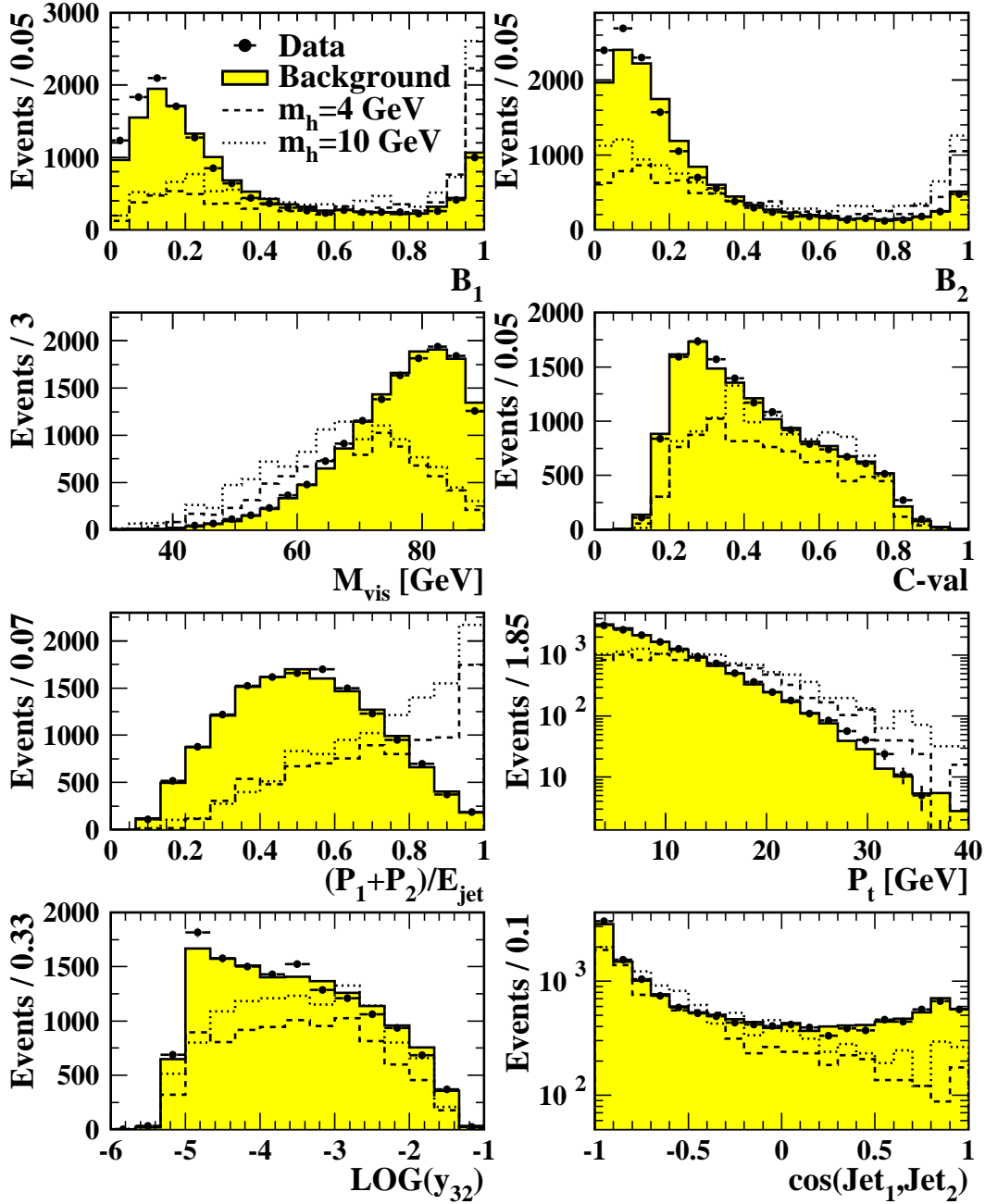}}
    \caption{{\it Distributions used in the likelihood selection for 
        preselected events and two hypothetical \ho\ Higgs masses. The
        points with error bars are data. The solid line is simulated
        background normalised to the recorded data. The dashed (dotted)
        line is a simulated Higgs boson \ho\ at a mass $m_{{\rm
            Higgs}}=4$ {\rm GeV} ($m_{{\rm Higgs}}=10$ {\rm GeV}) scaled
        arbitrarily for better visibility.}}
\label{LHVars}
  \end{center}
\end{figure}

From reference histograms of eight variables, listed below, we define
signal likelihood selections for each mass hypothesis, both for the
\ho\ and the \Ao\ Higgs boson.  Due to the overwhelming dominance of
the \qq\ background, we use a single inclusive background class.  The
eight reference histograms are (see Figure \ref{LHVars}):

\begin{enumerate}
\item $B_1$. The `b-ness' of the jet with the highest 
multiplicity. This value is defined as
\begin{equation}
B_1=\frac{L_b}{L_b+L_c+L_{uds}}.
\end{equation}
Here $L_b$, $L_c$ and $L_{uds}$ are likelihood values for bottom, charm 
and light flavour jets respectively~\cite{OPALPaper}.
\item $B_2$. The `b-ness' value of the jet with the second highest 
multiplicity.
\item $M_{vis}$. The measured invariant mass of the event.
\item The $C$-value of the event~\cite{CVAR}.
\item $(P_1+P_2)/E_{jet}$ in Jet(3). The sum of the momenta of the two
tracks in Jet(3) with the highest momentum divided by the measured
energy in Jet(3).
\item $P_t$. The transverse momentum of the event, with respect to the
beam axis.

\item $\log(y_{32})$. The logarithm of the $y_{32}$ value of the
event~\cite{Y43}.
\item $\cos(\mathrm{Jet(1)},\mathrm{Jet(2)})$. Cosine of the angle
  between the jet with the highest multiplicity (Jet(1)) and the one
  with the second highest multiplicity (Jet(2)).
\end{enumerate}

\begin{figure}[htb]
  \begin{center}
    \resizebox{180mm}{!}{
      \includegraphics{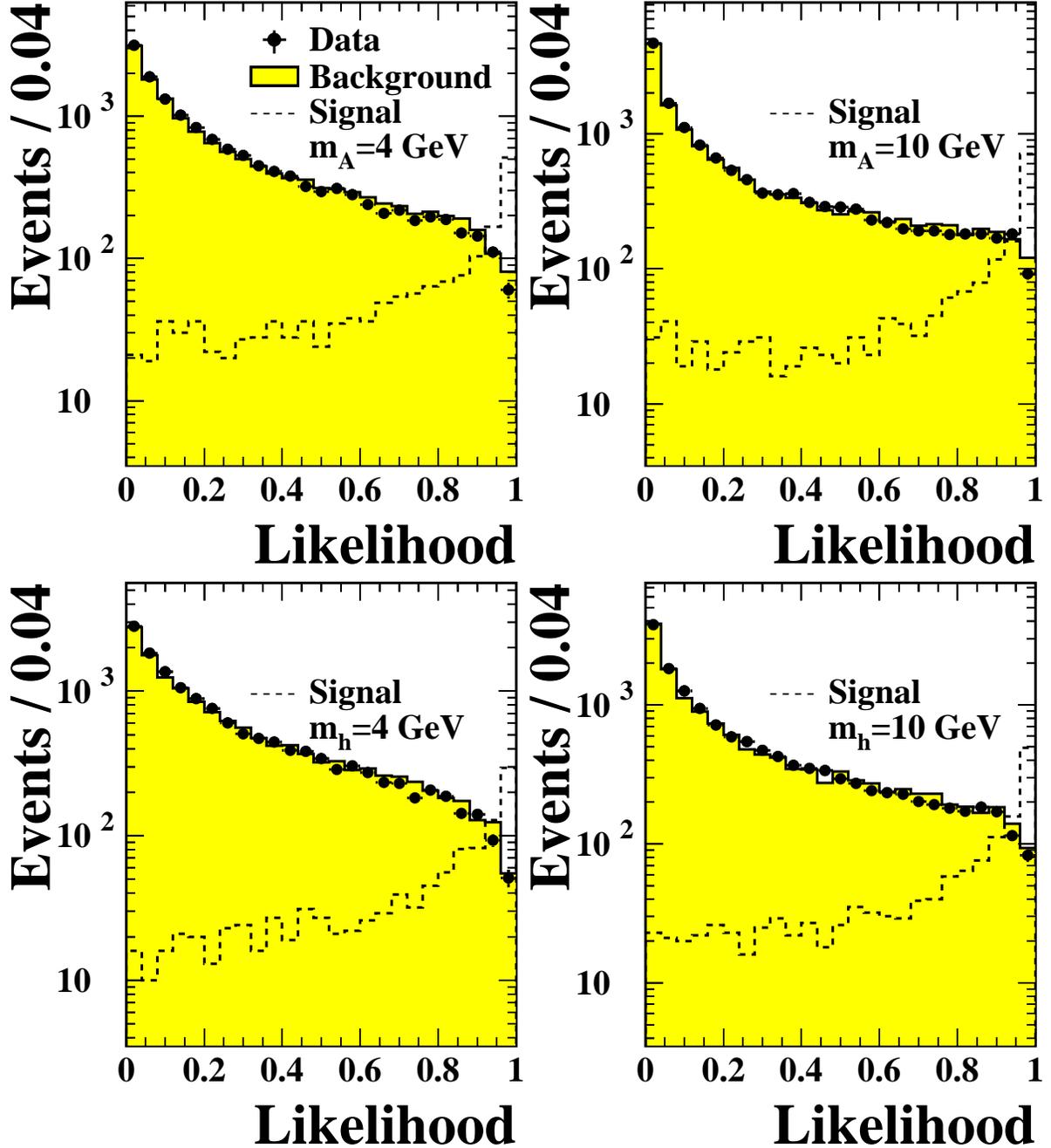}}
    \caption{{\it Likelihood distributions for the selection of a 
        CP-odd Higgs \Ao\ with masses of 4 and 10 {\rm GeV} and for
        a CP-even Higgs \ho\ with masses of 4 and 10 {\rm GeV}. The points with
        error bars are data. The solid line is simulated background 
        normalised to the recorded luminosity. The dashed line is a simulated
        Higgs boson scaled arbitrarily for better visibility.}}
    \label{LHDis1}
  \end{center}
\end{figure}

The likelihood distributions are shown in Figure \ref{LHDis1}.  The
likelihood cuts are determined separately for each Higgs type and mass
hypothesis in a compromise to achieve a good expected limit,
calculated with Monte Carlo experiments, and a smooth behavior of
efficiency and expected number of backgrounds as a function of the
Higgs mass. 
After the likelihood cut the data are in good agreement with background
Monte Carlo simulation (see Table~\ref{totresult_a} and 
Table~\ref{totresult_h}).

\begin{table}[htb]
\begin{center}
\begin{tabular}[ht]{|r|r||r||r|r|} \hline
Mass(\Ao ) & LH cut &Data & Total background & Efficiency\\ 
 (GeV)&  & Events & Events &  \% \\ \hline \hline
4 & 0.985 & 12 & 14.9$\pm$2.9$\pm$1.6 & 3.4$\pm$0.2$\pm$0.1 \\ \hline
5 & 0.985 & 17 & 16.0$\pm$3.1$\pm$1.7 & 3.5$\pm$0.2$\pm$0.1 \\ \hline
6 & 0.985 & 13 & 16.6$\pm$3.1$\pm$1.7 & 3.7$\pm$0.2$\pm$0.1 \\ \hline
7 & 0.987 & 14 & 13.8$\pm$2.8$\pm$1.4 & 3.8$\pm$0.2$\pm$0.2 \\ \hline
8 & 0.990 & 13 & 17.8$\pm$3.3$\pm$1.9 & 3.7$\pm$0.2$\pm$0.1 \\ \hline
9 & 0.990 & 11 & 15.1$\pm$3.0$\pm$1.6 & 3.9$\pm$0.2$\pm$0.2 \\ \hline
10& 0.990 & 11 & 17.6$\pm$3.2$\pm$1.8 & 3.8$\pm$0.2$\pm$0.2 \\ \hline
11& 0.990 & 13 & 19.1$\pm$3.3$\pm$2.0 & 4.0$\pm$0.2$\pm$0.2 \\ \hline
12& 0.992 & 13 & 18.3$\pm$3.3$\pm$1.9 & 3.9$\pm$0.2$\pm$0.2 \\ \hline
\end{tabular} 
\caption{ {\it The number of selected \Ao\ candidate events after 
the likelihood cut.}}
\label{totresult_a} 
\end{center}
\end{table}

\begin{table}[htb]
\begin{center}
\begin{tabular}[ht]{|r|r||r||r|r|} \hline
Mass(\ho ) & LH cut & Data & Total background & Efficiency\\(GeV)&  & Events & Events &  \% \\ \hline \hline
4 & 0.965 & 38 & 46.9$\pm$5.1$\pm$4.8 & 2.8$\pm$0.2$\pm$0.1 \\ \hline
5 & 0.970 & 41 & 36.9$\pm$4.6$\pm$3.8 & 3.1$\pm$0.2$\pm$0.1 \\ \hline
6 & 0.975 & 26 & 29.0$\pm$4.1$\pm$3.0 & 2.8$\pm$0.2$\pm$0.1 \\ \hline
7 & 0.980 & 22 & 20.5$\pm$3.4$\pm$2.1 & 3.2$\pm$0.2$\pm$0.1 \\ \hline
8 & 0.985 & 18 & 12.7$\pm$2.7$\pm$1.3 & 3.1$\pm$0.2$\pm$0.1 \\ \hline
9 & 0.987 & 16 & 15.3$\pm$3.0$\pm$1.6 & 3.2$\pm$0.2$\pm$0.1 \\ \hline
10& 0.985 & 15 & 17.9$\pm$3.2$\pm$1.8 & 3.0$\pm$0.2$\pm$0.1 \\ \hline
11& 0.990 & 11 & 10.1$\pm$2.4$\pm$1.0 & 3.3$\pm$0.2$\pm$0.1 \\ \hline
12& 0.993 &  9 &  9.9$\pm$2.4$\pm$1.0 & 3.3$\pm$0.2$\pm$0.1 \\ \hline
\end{tabular}
\caption{ {\it The number of selected \ho\ candidate events after 
the likelihood cut.}}
\label{totresult_h} 
\end{center}
\end{table}

\section{Systematic Studies}
In order to estimate systematic uncertainties, the following sources of
uncertainty are varied in the Monte Carlo, one at a time, and the
analysis is redone with the adjusted samples to calculate the
difference in the background expectation and the efficiency with respect to
the standard analysis.  For variations done in two directions, with
respect to the central value, the largest deviation in expected
background and efficiency after re-analysis is taken as the systematic
uncertainty.  The following sources of uncertainty were studied:
\begin{itemize}

\item The systematic error contributed by the b-tagging due to track resolution
  modelling. This was estimated with a variation of the track parameters 
  $\phi_0\pm 5$\%, $d_0\pm 5$\%, $z_0 \pm 10$\% as described in 
  \cite{Higgs183}. This variation  particularly influences the values of 
  track- and secondary vertex significances ($d/\sigma_d$ and $l/\sigma_l$)
  which are important variables for the b-tagging procedure. 

\item The uncertainty in the b-quark fragmentation
function~\cite{Peterson} is estimated by varying $\epsilon_{\rm b}$ by
$\pm 25$\% around a central value of $\epsilon_{\rm
b}=0.0038$~\cite{HFWG}.  A smaller(higher) value of $\epsilon_{\rm b}$
corresponds to a harder(softer) B-hadron spectrum.

\item Description of the kinematic variables used in the likelihood selection.
  The kinematic likelihood variables in the Monte Carlo were 
  shifted to match the
  mean of the data. After the shift on the variable the selection
  is reapplied and the deviations in the expected background 
  and efficiency are taken as systematic errors.

\item The B-hadron charged decay multiplicity uncertainty~\cite{bmul}.

\end{itemize}

The systematic deviations are calculated separately for the two
different hypotheses on Higgs type (\Ao\ and \ho) after a likelihood
cut at 0.8 to ensure that statistical contributions to the estimated
systematic errors are minimal.  

Furthermore the analysis for one typical Higgs hypothesis (\mh$=$5
GeV) was redone under the assumption of 100\%\ correlated helicity
states of the $\tau$'s. The deviation of 1.3\%\ for signal and 1.5\%\ 
for background expectation were included as a source of systematic
uncertainty. The contributions to the systematic uncertainty are
broken down for the analyses in Table \ref{systab}.

Adding in quadrature the statistical uncertainties and the
uncertainties from the above sources yields the total errors, listed
in Table \ref{totresult_a} and Table \ref{totresult_h}, on the
selection efficiency and background rates for all nine masses of \Ao\ and
\ho.  The uncertainty due to the b-fragmentation function is the
highest contribution.

\begin{table}
  \begin{center}
    \begin{tabular}{|c||c|c|c|c|} \hline
       &\multicolumn{2}{|c|}{\Ao\ Selection}&
        \multicolumn{2}{|c|}{\ho\ Selection} \\
       & Signal eff.  & \qq\ background & Signal eff.  & \qq\ background\\
      Variation of & \% & Events & \% & Events \\ \hline \hline
      Track parameters & 0.2 & 4.1 & 2.4 & 1.7 \\ 
      $\epsilon_{b}$   & 2.7 & 8.3 & 1.7 & 9.2 \\
      $B$ multiplicity & 1.8 & 1.8 & 1.8 & 1.8 \\
      $M_{vis}$        & 0.3 & 2.6 & 0.9 & 1.8 \\
      $(P_1+P_2)/E$    & 0.3 & 0.3 & 0.1 & 0.2 \\
      $C$-value        & 0.6 & 3.4 & 0.8 & 2.7 \\ 
      $P_t$            & 0.1 & 0.6 & 0.1 & 0.5 \\
      $\log(y_{32})$   & 0.5 & 0.8 & 0.2 & 0.2 \\
      $\cos({\rm Jet(1)},{\rm Jet(2)})$ & 1.7 & 1.1 & 0.3 & 1.1 \\ 
      $\tau$ correlations & 1.3 & 1.5 & 1.3 & 1.5 \\ \hline
      Total  & 4.0 & 10.5 & 3.9 & 10.3 \\ \hline
    \end{tabular}
    \caption{{\it Relative change of the signal efficiency and number
        of expected background events for the selection of \Ao\ and
        \ho\ at a likelihood cut, $L > 0.8$. The total
        value expresses the quadratically added contributions.}}
        \label{systab} \end{center}
\end{table}

\section{Limits on $\fxah$ in the 2HDM Type II}
The results of the selection with systematic and statistical errors
are listed in Table \ref{totresult_a} for the \Ao\ and in Table
\ref{totresult_h} for the \ho.

2HDM Type II limits are determined for the cross-section of the
process \epm $\rightarrow$ \Zo $\rightarrow$ \bb\tptm.  Due to the
dependence of the cross-section (Eqn. 4) on the enhancement factor
$\xahd$, a limit on $\xi$ is obtained by calculating
$N_{95}$, the 95\% C.L. upper limit on the rate of accepted signal 
events in the data, according to~\cite{PDG96}, and adjusting for the
efficiency, cross section, and luminosity.

The total error on the efficiency and on the background estimation are
convoluted into the limit according to \cite{CousinFeldman}. For the
mass points between the measured ones, in one GeV steps from 4 to
12 GeV, the limit is linearly interpolated from the two neighbouring
measurements.  
Comparing limits obtained with analyses optimized to neighbouring mass
points, the uncertainty of the interpolation was estimated to be less
than 0.5 units in $\xi$.

\begin{table}[h]
\begin{center}
\begin{tabular}{|c|c||c|c|c|c|c|c|c|c|c|} \hline
Type & \mah\ GeV & 4 & 5 & 6 & 7 & 8 & 9 & 10 & 11 & 12 \\ \hline \hline
\Ao\ &expected $\xnfa$ & 9.5&10.3&11.0&11.1&12.9&13.4&14.4&15.5&16.6\\ \hline
\Ao\ &observed $\xnfa$ & 8.5&11.0& 9.6&11.5&10.7&11.0&11.3&12.3&13.6\\ \hline \hline
\ho\ &expected  $\xnfh$ &10.2&10.3&11.3&10.8&10.5&12.1&13.7&11.9&12.9\\ \hline
\ho\ &observed $\xnfh$ & 8.2&11.8&10.4&11.8&13.7&12.6&12.6&12.7&12.9\\ \hline
\end{tabular}
\caption{{\it The upper limit on $\xi$ in a 2HDM Type II model for
masses $m_{Higgs}=4$ {\rm GeV} to $m_{Higgs}=12$ {\rm GeV} at
95\%\ C.L. calculated assuming 100\% branching ratio of the \Ao\ (upper
part) and \ho\ (lower part) into \tptm .}}
\label{newlimits-a}
\end{center}
\end{table}

\begin{figure}[htb] 
  \begin{center} \resizebox{170mm}{!}{
    \includegraphics{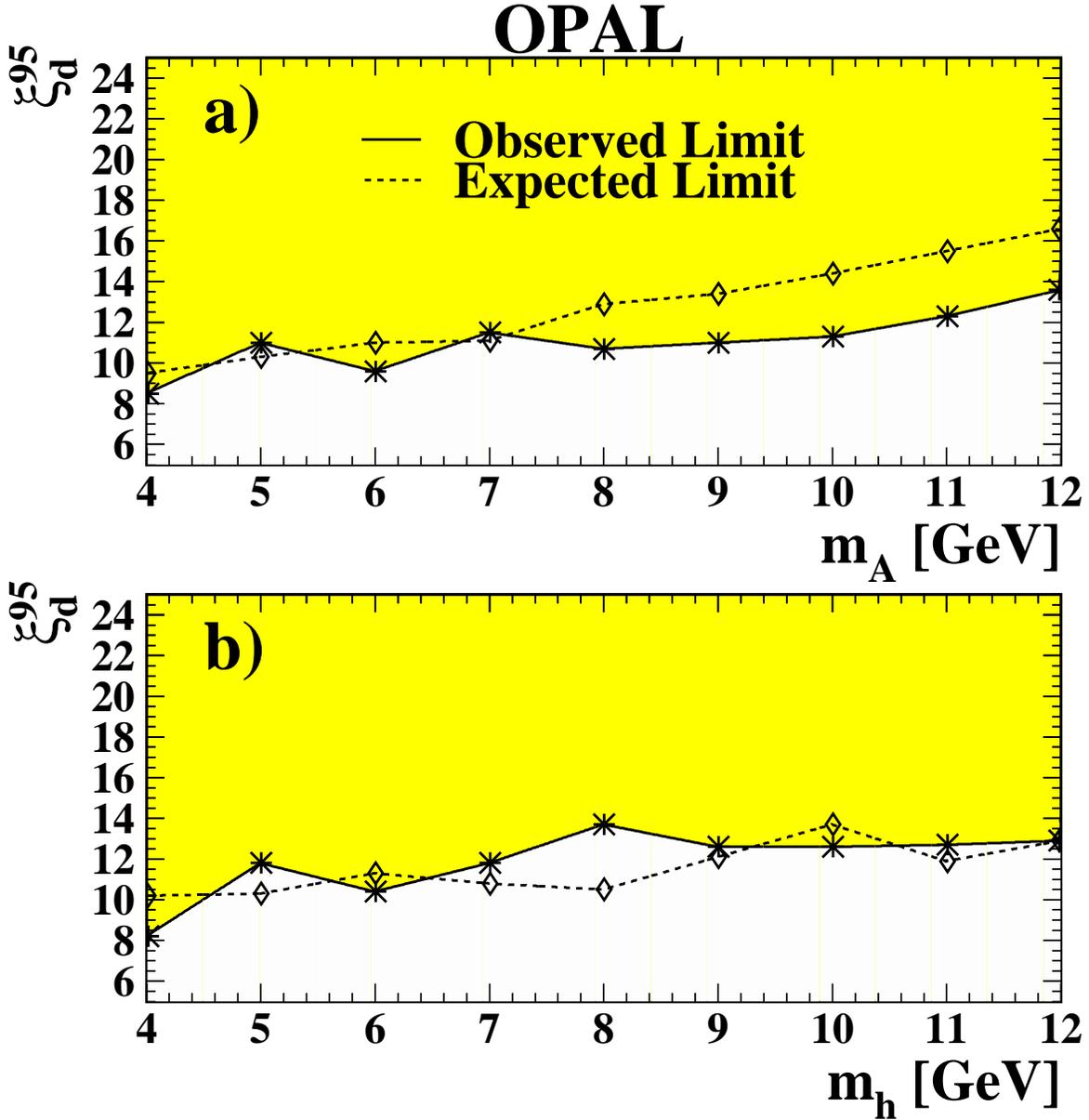}} \caption{{\it Excluded values
    of \xid\ at 95\% C.L. (dark grey region) in the 2HDM type II model for
    the Yukawa production of a CP-odd \Ao\ (upper plot) and for
    the CP-even \ho\ (lower plot) assuming the branching fraction of
    the Higgs boson into \tptm to be 100\%.  The expected (diamonds)
    and observed (stars) limits have been calculated at specific mass
    points (4--12 {\rm GeV} in one {\rm GeV} steps) and linearly
    interpolated in between. }} \label{limits} \end{center}
\end{figure}

Assuming a branching ratio of 100\% for Higgs boson decays into \tptm\ 
the limits on $\xahd$ are shown
in Figure \ref{limits}(a) for \Ao\ production and in
Figure \ref{limits}(b) for \ho\ production, and are 
summarised in Table \ref{newlimits-a}. 

In a 2HDM model with Standard Model particle content, the Higgs
branching ratio into \tptm\ for $\xi_d \approx 10$ is about 85\% for
Higgs masses between 4 and 9.4 GeV. In the mass range from 9.4~Gev to
11.0~GeV the branching ratios are very much influenced by mixing of
the Higgs bosons \ho\ and \Ao\ with \bb\ bound states with the same
quantum numbers (see Table~\ref{mixingpartnersa}).  We have therefore
calculated the branching ratios of the Higgs into \tptm\ according to
reference \cite{DreesMixing}.  The limits derived within this model
are shown in Figure \ref{limits2_lin}(a) and (b) for CP-odd and
CP-even Higgs production, respectively. The observed structure, in
particular at higher Higgs boson masses, is a consequence of the
behaviour of the branching ratio in this particular model.

\begin{table}[htb]
\begin{center}
\begin{tabular}{|r||r|r|} \hline
state & $\eta$ Mass & $\chi_0$ Mass \\ 
    $n$       & GeV & GeV \\ \hline
1 &  9.412 &  9.860 \\
2 &  9.992 & 10.235 \\
3 & 10.340 & \\
4 & 10.570 &\\
5 & 10.846 &\\
6 & 11.014 &\\ \hline
\end{tabular}
\caption{\it The mass of the $\eta$ states assumed to mix with the pseudoscalar
\Ao\ and of the $\chi_0$ states assumed to mix with the scalar \ho, taken from
\cite{DreesMixing}.}
\label{mixingpartnersa}
\end{center}
\end{table}

\section{Implications for the Muon Anomalous Magnetic Moment}
Recent measurements of the anomalous magnetic moment,
$a_{\mu}=\frac{1}{2}(g-2)_{\mu}$, of the muon have given a result
which deviates from the Standard Model expectation by $\approx
400\times 10^{-11}$, corresponding to about 2.6 standard
deviations~\cite{gmin2}. Depending on the estimation of the hadronic
contribution to the muon anomalous magnetic moment, the 90\%\ C.L.
ranges for the contribution of New Physics $\delta a_{\mu}({\rm NP})$
are:
\begin{eqnarray}
215\times10^{-11} \le & \delta a_{\mu}({\rm NP}) & \le 637\times10^{-11}~\cite{Marciano}\\
170\times10^{-11} \le & \delta a_{\mu}({\rm NP}) & \le 690\times10^{-11}~\cite{Hocker}\\
112\times10^{-11} \le & \delta a_{\mu}({\rm NP}) & \le 573\times10^{-11}~\cite{Narison}
\end{eqnarray}

\begin{figure}[htb] 
  \begin{center}
    \resizebox{170mm}{!}{   
      \includegraphics{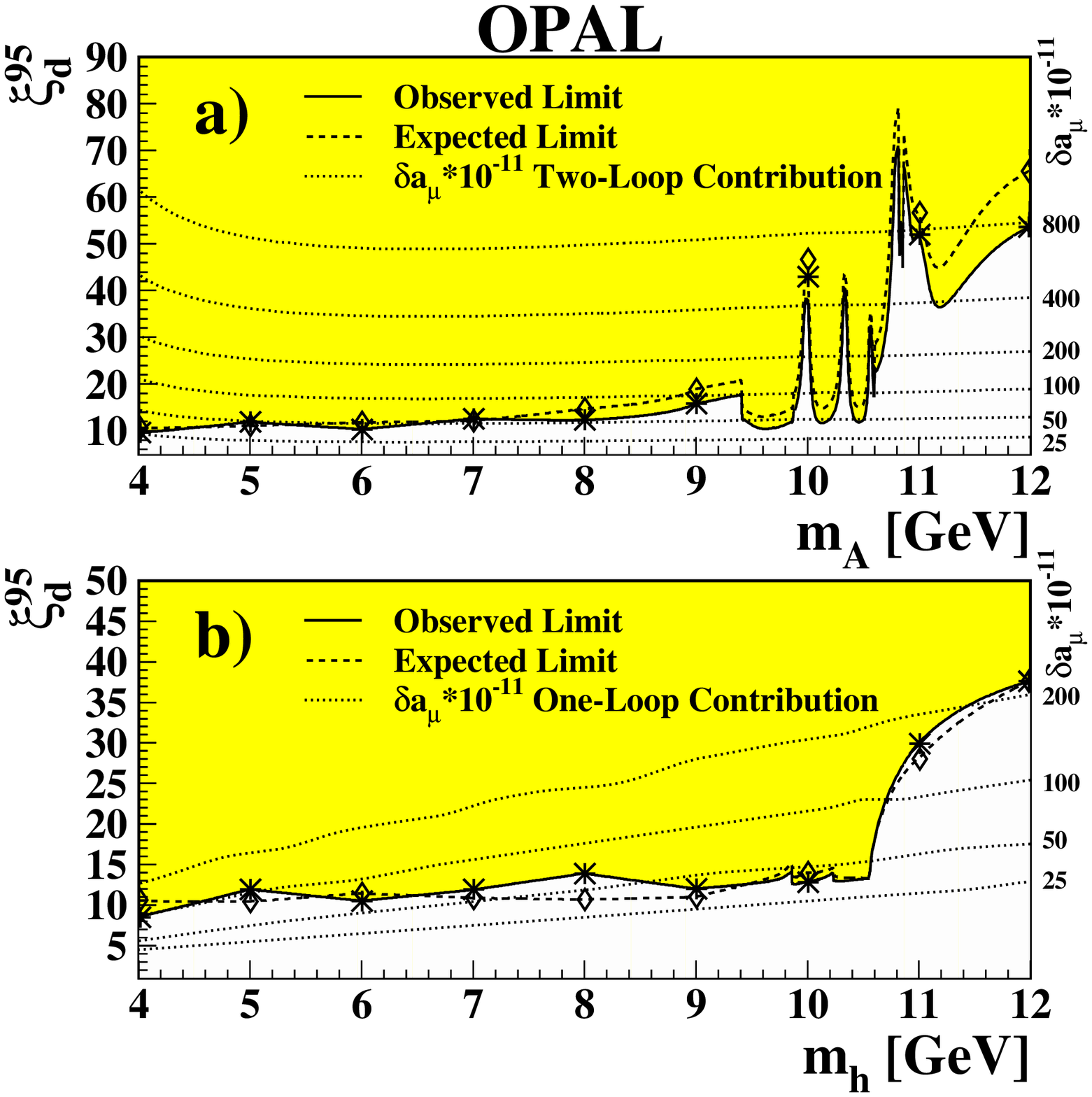}}

    \caption{{\it Excluded values of \xid\ at 95\% C.L. in the 2HDM
        type II model with Standard Model particle content for the
        Yukawa production of a CP-odd \Ao\ (upper plot) and for the
        CP-even \ho\ (lower plot) with the mixing to \bb\ bound states
        taken into account.  The structure results entirely from the
        theoretically suggested branching ratios \cite{DreesMixing}.
        The expected (diamonds) and observed (stars) limits have been
        calculated at specific mass points (4--12 {\rm GeV} in one
        {\rm GeV} steps) and linearly interpolated in between. The
        dotted lines are the contours of the predicted Higgs 
        contribution (one-loop~\cite{oneloop,Dedes} for the
        CP-even and two-loop~\cite{twoloop,Mariag2} for the
        CP-odd Higgs) to the muon anomalous magnetic moment, $\delta
        a_\mu$(Higgs) (in units of $10^{-11}$).
        }}  \label{limits2_lin}
        \end{center}
\end{figure}

Light Higgs bosons A and h could form a part of $a_\mu$ via loop
diagrams. A one-loop calculation~\cite{oneloop} predicts positive
contributions $\delta a_{\mu}^{I}$(h) $>0$ for the h, and negative
contributions $\delta a_{\mu}^{I}$(A) $<0$ for the A.  The two-loop
terms, due to the stronger coupling of the Higgs fields to loops
with heavy quarks, turn out to be larger in magnitude than the
one-loop terms, and of opposite sign~\cite{twoloop}, giving a total
positive contribution $\delta a_{\mu}^{II}$(A) $>0$ for the \Ao , as
shown with indicated isolines in Figure~\ref{limits2_lin}
(a)~\cite{Mariag2}. However, the two-loop terms gives a total negative
contribution $\delta a_{\mu}^{II}$(h) $<0$ for the \ho , thus
suggesting that the \ho\ can not account for the BNL observation.  We
show in Figure~\ref{limits2_lin} (b) only the isolines of the
contribution from the earlier one-loop calculation~\cite{Dedes} which
resulted in a positive value of $\delta a_{\mu}^{I}$(h). Our data
exclude positive contributions $\delta a_{\mu}^{I}$(h) $>
100\times10^{-11}$ for h masses between 4.0 and 10.7~GeV at the
one-loop level, and $\delta a_{\mu}^{II}$(A) $> 100\times10^{-11}$ for
A masses betwen 4.0 and 9.9~GeV at the two-loop level.  Similar limits
have been derived from radiative $\Upsilon$ decays~\cite{Mariag2,CUSB}
for h/A masses lighter than about 8 GeV, with, however, large QCD
uncertainies.

In reference~\cite{Dedes} and \cite{Mariag2} the authors have
suggested that a light Higgs boson could fully account for the
observed deviation of the measured $(g-2)_{\mu}$ from the Standard
Model expectation. In a scenario without contributions of other new
particles, eg. gauginos, and assuming that either \ho\ or \Ao\ is
heavy enough to render associated \Ao\ho\ production inaccessible at
LEP, only the lighter of the two Higgs bosons would sizeably
contribute. For a light h, one has to assume in addition
$\sin(\beta-\alpha)\approx 0$ to explain its non-observation in the
Standard Model search for the Higgsstrahlung process.
The experimental results of this analysis can be interpreted in such a
scenario and would rule out (using the one-loop
calculation~\cite{Dedes}) a light h in the mass range of 4--10.7~GeV,
and (using the two-loop calculation~\cite{twoloop}) a light A in the
mass range from 4--9.9~GeV as the only source of the discrepancy
in the $(g-2)_{\mu}$ measurement for all three 90\%\ C.L.  ranges listed
above.

\section{Conclusion}
The Yukawa production of light neutral Higgs bosons in the channel
\epm $\rightarrow$ \bb\Ao/\ho $\rightarrow$ \bb\tptm\ is studied.
The search presented here, based on data collected by OPAL at
$\sqrt{s} \approx \mZ$ in the years 1992 to 1995, has not revealed any
significant excess over the expected background.  Limits on the Yukawa
production of a light Higgs with masses in the range of 4 GeV to 12
GeV have been set at 95\% C.L.  New limits on the parameters
$\xad=|\tan\beta|$ and $\xhd=|\sin\alpha/\cos\beta|$ are presented for
\Ao\ and \ho\ production, respectively. Assuming a branching ratio
to \tptm\ of 100\% , upper limits for $\xad$ can be set within the range of 8.5
to 13.6 and for $\xhd$ between 8.2 to 13.7, depending on the mass of
the Higgs boson.  In a 2HDM type II model with Standard Model particle
content similar limits are obtained up to masses of 9.4~GeV. Above
9.4~GeV the mixing of the Higgs bosons \ho\ and \Ao\ with \bb\ bound
states with the same quantum numbers, especially for the
CP-odd \Ao, results in weaker limits in certain mass ranges.
The experimental result of this analysis restricts the contribution
to the anomalous magnetic moment of the muon for a light h in the mass
range of 4--10.7~GeV (using one-loop calculation~\cite{oneloop}) and
for a light A in the mass range from 4--9.9~GeV (using the two-loop
calculation of~\cite{twoloop}) to $\delta a_{\mu}({\rm Higgs}) <
100\times10^{-11}$ at the 95\% C.L.

\section{Acknowledgements}
We particularly wish to thank the SL Division for the efficient
operation of the LEP accelerator at all energies and for their close
cooperation with our experimental group.  We thank our colleagues from
CEA, DAPNIA/SPP, CE-Saclay for their efforts over the years on the
time-of-flight and trigger systems which we continue to use. We also
gratefully thank A.~Dedes and M.~Krawczyk for valuable support on
theoretical issues.  In addition to the support staff at our own
institutions we are pleased to acknowledge the \\ Department of
Energy, USA, \\ National Science Foundation, USA, \\ Particle Physics
and Astronomy Research Council, UK, \\ Natural Sciences and
Engineering Research Council, Canada, \\ Israel Science Foundation,
administered by the Israel Academy of Science and Humanities, \\
Minerva Gesellschaft, \\ Benoziyo Center for High Energy Physics,\\
Japanese Ministry of Education, Science and Culture (the Monbusho) and
a grant under the Monbusho International Science Research Program,\\
Japanese Society for the Promotion of Science (JSPS),\\ German Israeli
Bi-national Science Foundation (GIF), \\ Bundesministerium f\"ur
Bildung und Forschung, Germany, \\ National Research Council of
Canada, \\ Research Corporation, USA,\\ Hungarian Foundation for
Scientific Research, OTKA T-029328, T023793 and OTKA F-023259.\\

\cleardoublepage

\cleardoublepage

\begin{thebibliography}{99}

\bibitem{LEPSM}
ALEPH, DELPHI, L3, OPAL Collaborations and the LEP working group for
Higgs boson searches, {\em Search for the Standard Model Higgs Boson
at LEP}, CERN-EP/2001-055 (2001).

\bibitem{Where}
M.\ Krawczyk, \APP{29}{1998}{3543}.

\bibitem{CKZ}
P.H.~Chankowski, M.~Krawczyk~\etal, Eur. Phys. J {\bf C11} (1999) 661. 

\bibitem{Hunter}
J.\ Gunion~\etal, {\em The Higgs Hunter's Guide}, Addison-Wesley (1990).

\bibitem{LEPHiggs}
ALEPH, DELPHI, L3, OPAL Collaborations and the LEP working group for
Higgs boson searches, {\em Searches for Higgs Bosons: Preliminary Combined 
Results Using LEP Data Collected at Energies Up To 202 GeV}, 
CERN-EP/2000-055 (2000) 28.

\bibitem{LimitAB}
The OPAL Collaboration, G. Abbiendi~\etal, Eur. Phys. J. {\bf C18} (2001) 425.

\bibitem{OPALPaper}
The OPAL Collaboration, G. Abbiendi~\etal, Eur. Phys. J. {\bf C12} (2000) 567.

\bibitem{Maria1}
J. Kalinowski and M. Krawczyk, \APP{27}{1996}{961}.

\bibitem{Durham}
Y.\ Dokshitzer, J. Phys. {\bf G17} (1991) 1537.

\bibitem{OPALDetector}
The OPAL Collaboration, K. Ahmet~\etal, Nucl. Instrum. Meth. {\bf A305} 
(1991) 275.

\bibitem{OPALZ1}
The OPAL Collaboration, G. Alexander~\etal, \ZPC{52}{1991}{175}.

\bibitem{Lumi}
The OPAL Collaboration, G. Abbiendi~\etal, Eur. Phys. J. {\bf C19} (2001) 587.

\bibitem{OPALZ2}
The OPAL Collaboration, R. Akers~\etal, \ZPC{65}{1995}{17}.

\bibitem{osim}
The OPAL Collaboration, J. Allison~\etal, Nucl. Instrum. Meth. {\bf A317} 
(1992) 47.

\bibitem{Vermaseren}
J.A.M.~Vermaseren, Nucl. Phys. {\bf B229} (1983) 347.

\bibitem{PHOJET} R.~Engel, Z. Phys. {\bf C66} (1995) 203; \\
R.~Engel and J.~Ranft, Phys. Rev. {\bf D54} (1996) 4244.

\bibitem{Fermisv}
J.\ Hilgart, R.\ Kleiss and F.\ Le Diberder, \CPC{75} (1993) 191.

 \bibitem{grc4f} 
J. Fujimoto et al., Comp. Phys. Comm. {\bf 100} (1997) 128.

\bibitem{Jetset}
T.\ Sj{\"o}strand, {\em High-energy-physics event generation with PYTHIA 5.7
 and JETSET 7.4}, \CPC{82} (1994) 74.

\bibitem{OPMOD}
The OPAL Collaboration, G. Alexander~\etal, \ZPC{69}{1996}{543};\\
The OPAL Collaboration, J. Allison~\etal, Nucl. Instrum. Meth. {\bf A317} 
(1992) 47.

\bibitem{Tauola}
S. Jadach, Z. Was~\etal, \CPC{76} (1993) 361.

\bibitem{MT}
The OPAL Collaboration, K. Ackerstaff~\etal, Eur. Phys. J. {\bf C2} (1998) 213.

\bibitem{CVAR}
G. Paris, \PLB{74}{1978}{65}.

\bibitem{Y43}
S. Catani \etal, \PLB{269}{1991}{432}.

\bibitem{elecid}
The OPAL collaboration, G.~Alexander~\etal, \ZPC{70}{1996}{357}.

\bibitem{muonid}
The OPAL Collaboration, M.Z.~Akrawy~\etal, \PLB{263}{1991}{311}.

\bibitem{Higgs183}
The OPAL Collaboration, G. Abbiendi~\etal, Eur. Phys. J. {\bf C12} (2000) 567.

\bibitem{Peterson}
C.\ Peterson~\etal, \PhysRev\ {\bf D27} (1983) 105. 

\bibitem{HFWG}
ALEPH, DELPHI, L3, OPAL, CDF and SLD Collaborations, 
{\em Combined results on b-hadron production rates, lifetimes, oscillations and semileptonic decays}, 
CERN-EP/2000-096 (2000).

\bibitem{bmul}
The OPAL Collaboration, R. Akers~\etal, \ZPC{61}{1994}{209}.

\bibitem{PDG96}
Particle Data Group, \PRD{54}{1996}{1}.

\bibitem{CousinFeldman}
R.\ Cousins and V.\ Highland, \NIMA{320} (1992) 331.

\bibitem{DreesMixing}
M. Drees and K.-I. Hikasa, \PRD{41}{1990}{1547}.

\bibitem{gmin2}
Muon (g-2) Collaboration, H.N. Brown~\etal, Phys. Rev. Lett. {\bf 86} 
(2001) 2227.

\bibitem{Marciano}
A.~Czarnecki and W.J.~Marciano, Phys. Rev. {\bf D64} (2001) 013014. 

\bibitem{Hocker}
M. Davier and A. H\"ocker, Phys. Lett. {\bf B435} (1998) 427.

\bibitem{Narison}
S. Narison, {\em Muon and tau anomalies updated}, hep-ph/0103199v3 (2001).

\bibitem{oneloop}
J.P.~Leveille, Nucl. Phys. {\bf B137} (1978) 63.

\bibitem{twoloop}
D. Chang, W.-F. Chang~\etal, Phys. Rev. {\bf D63} (2001) 091301;
K. Cheung, C.-H. Chou~\etal, {\em Muon Anomalous Magnetic Moment, 
Two-Higgs-Doublet Model, and Supersymmetry}, hep-ph/0103183.

\bibitem{Mariag2}
M. Krawczyk, {\em The new $(g-2)_\mu$ measurement and limit on the 
light Higgs bosons in 2HDM (II)}, hep-ph/0103223v3 (Submitted to
Phys. Rev. {\bf D}). 

\bibitem{Dedes}
A. Dedes and H.E. Haber, JHEP {\bf 0105} (2001) 6.

\bibitem{CUSB}
CUSB Collaboration, P.~Franzini~\etal, Phys. Rev. {\bf D35} (1987) 2883.

\end{thebibliography}
\end{document}